# Tunable Magnetic Transition to a Singlet Ground State in a 2D Van der Waals Layered Trimerized Kagomé Magnet

Christopher M. Pasco[1,2], Ismail El Baggari[3], Elisabeth Bianco[4], Lena F. Kourkoutis[4,5], Tyrel M. McQueen[1,2,6,*]

[1]Department of Chemistry, The Johns Hopkins University, Baltimore, MD 21218, United States

[2]Institute for Quantum Matter, Department of Physics and Astronomy, The Johns Hopkins University, Baltimore, MD 21218, United States

[3]Department of Physics, Cornell University, Ithaca, NY 14853, United States

[4]Kavli Institute at Cornell for Nanoscale Science, Cornell University, Ithaca, NY 14853, United States

[5]School of Applied and Engineering Physics, Cornell University, Ithaca, NY 14853, United States

[6]Department of Materials Science and Engineering, The Johns Hopkins University, Baltimore, MD 21218, United States

*Corresponding author: mcqueen@jhu.edu

**ABSTRACT:** Incorporating magnetism into two dimensional (2D) van der Waals (VdW) heterostrutures is crucial for the development of functional electronic and magnetic devices. Here we show that $Nb_3X_8$ (X = Cl, Br) is a family of 2D layered trimerized kagomé magnets that are paramagnetic at high temperatures and undergo a first order phase transition on cooling to a singlet magnetic state. X-ray diffraction shows that a rearrangement of the VdW stacking accompanies the magnetic transition, with high and low temperature phases consistent with STEM images of the end members α-$Nb_3Cl_8$ and β-$Nb_3Br_8$. The temperature of this transition is systematically varied across the solid solution $Nb_3Cl_{8-x}Br_x$ (x = 0-8), with x = 6 having transitions near room temperature. The solid solution also varies the optical properties, which are further modulated by the phase transition. As such, they provide a platform on which to understand and exploit the interplay between dimensionality, magnetism, and optoelectronic behavior in VdW materials.

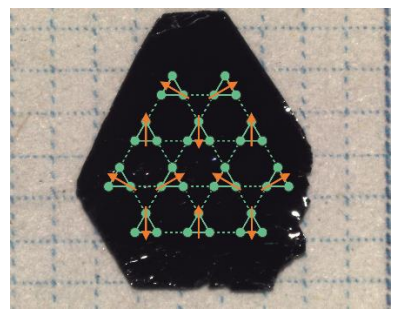

The discovery of graphene has recently sparked a surge of interest in other two dimensional (2D) semiconductors, such as $MoS_2$ and other transition metal dichalcogenides (TMDs), which exhibit a variety of properties that can be exploited in the production of devices.[1] One feature that makes these materials particularly desirable is that their layers are stacked by Van der Waals (VdW) forces which can make exfoliation relatively trivial. Variations in the basic properties of the atoms used within the family of TMDs have led to a range of potential applications.[1] Generically, TMDs share similar band structures, a lack of magnetic behaviors, and complex phenomena associated with quantum materials. One way to gain access to magnetic physics is to look at 2D semiconductors outside of the TMD family. One such potential variation is the trimerized kagomé magnet, in which metallic clusters of atoms carry a net magnetic moment and are embedded in a larger lattice with low dimensional exchange pathways between them.[2]

Geometrically frustrated magnetic materials are those in which magnetic units are embedded in a lattice in such a way that there is not a single lowest energy configuration of spins. When the interactions between these spins is strong, the inability to satisfy competing exchanges can, to avoid long range order, result in the formation of exotic ground states such as valence bond solids,[3] and spin liquids.[4] They can also drive significant structural distortions to break the frustrating symmetry such as dimerization in the spin-Peierls distortion.[5] Recently, significant interest has revolved around 2D magnetic materials, including single layer ferromagnets such as $CrI_3$[6] and $Fe_3TeGe_2$,[7] and the Kitaev spin liquid candidate α-$RuCl_3$.[8] Typically, the magnetic units are spins localized to individual ions, however, this is not true in the case of the trimerized kagomé magnet, where the free spins are delocalized across a metal-metal bonded cluster of ions, such as in $LiZn_2Mo_3O_8$.[9,10]

Originally reported in the 1960s,[11-13] the $Nb_3X_8$ family of materials hosts a trimerized kagomé lattice, in which each $Nb_3$ trimer is held together by strong metal-metal bonds, sharing a single unpaired electron, as seen in Figure 1. These effective S = ½ units are arranged in a geometrically frustrated triangular network. In 1992 it was reported that $Nb_3Cl_8$ undergoes a magnetic phase transition around T = 90 K.[14] At room temperature $Nb_3Cl_8$ behaves like a paramagnet, undergoing a first order phase transition to an apparently non-magnetic, *i.e.*, singlet, low temperature phase.[15,16] Individual layers are only weakly bound to each other by VdW interactions, and as a consequence these materials are also readily exfoliatable as seen in Figure 2.

In this article, we report the discovery that $Nb_3Br_8$ also undergoes a magnetic transition from a high temperature paramagnetic phase to a singlet phase above room temperature, at T = 382 K. In addition, for the continuous solid solution $Nb_3Cl_{8-x}Br_x$, the temperature of this phase transition can be tuned from T = 92 K to T = 387 K. The UV-visible optical absorption peaks are also systematically varied. Single crystal X-ray diffraction (SXRD) shows that the magnetic transition from the high temperature, paramagnetic state (α phase) to the low temperature, singlet state (β phase) is accompanied by a rearrangement of the stacking sequence between VdW layers of the native heterostructure. This is consistent with the results of atomic resolution scanning transmission electron microscopy (STEM) imaging of the end members α-$Nb_3Cl_8$ and β-$Nb_3Br_8$. Thus, the $Nb_3X_8$ family represents a group of 2D-exfoliatable materials with magnetic and optical properties tunable over a range that includes room temperature and visible light, making it potentially useful for the production of devices.

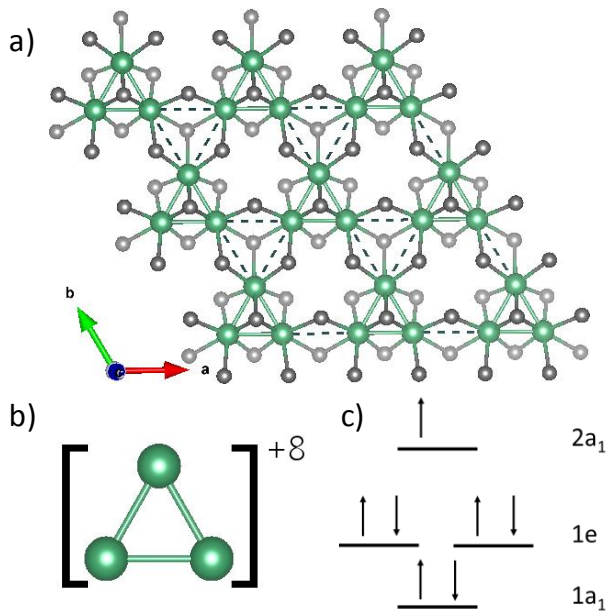

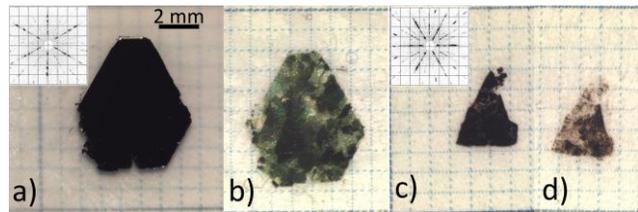

**Figure 2.** a) and c) are first exfoliations of $Nb_3Cl_8$ and $Nb_3Br_8$ crystals, respectively on a 1 mm grid, with backscatter Laue diffraction patterns of each in the (hk0) plane included as insets. b) and d) show the same samples after repeated exfoliations until they were thin enough to transmit light.

**Figure 1.** a) Single layer of $Nb_3X_8$ showing the trimerized kagomé structure of individual layers. The Nb atoms (green) are connected by a solid line showing the metal-metal bonds and a dashed grey line indicating unbonded triangles. Dark grey atoms are halides above the plane of $Nb_3$ clusters, and light grey atoms are halides below the plane. b) An individual $Nb_3$ cluster and its charge. c) Simplified molecular orbital diagram for the cluster showing a single unpaired electron.

## RESULTS AND DISCUSSION

**Structure (SXRD and STEM).** The structures of both α-$Nb_3Cl_8$ and β-$Nb_3Br_8$ are well known.[11,13] As referenced earlier, α-$Nb_3Cl_8$ is known to undergo a phase transition at T = 92 K to a non-magnetic β phase, however, the exact description of the structure of this low temperature phase has varied between reports. This is likely due to significant remnants of the high temperature phase as stacking faults after the transition.[15] For $Nb_3Br_8$ has been reported that an α variant isostructural to α-$Nb_3Cl_8$ was occasionally found in samples grown at T = 400 K, though it does not appear to have been extensively studied.[12] It is worth noting that this is very close to the temperature at which β-$Nb_3Br_8$ was found to transition into a high temperature phase in this work. This suggests that poor crystallinity due to the low temperature synthesis conditions may have trapped $Nb_3Br_8$ in its high temperature phase.

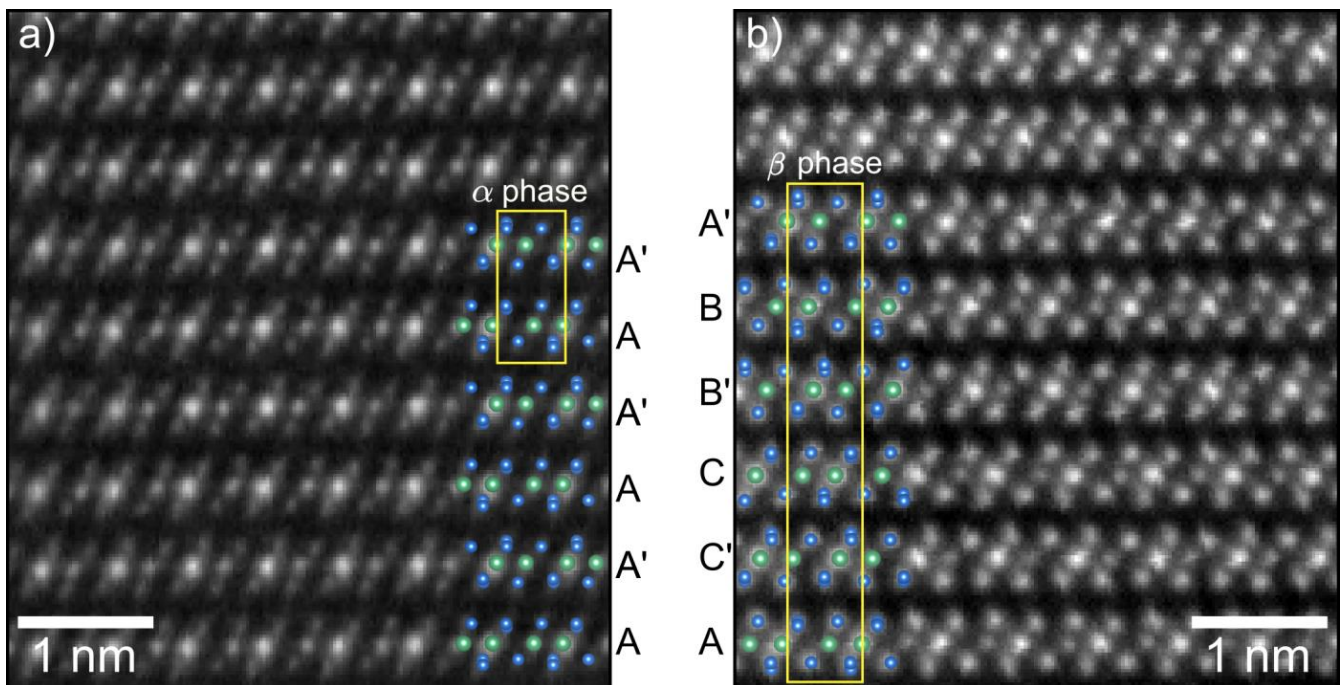

**Figure 3.** a) STEM image of α-$Nb_3Cl_8$ along (100) at T = 93 K with structural overlay. d) STEM image of β-$Nb_3Br_8$ along (100) showing the low temperature beta phase at T = 300 K with structural overlay. Unit cells are indicated by the yellow rectangles for each phase. In the structural overlays Nb atoms are shown in green and the halides in blue. Brighter spots in the Nb rows in the STEM images are where two Nb atoms in the unit cell are overlaid in this projection, while the dimmer spot comes from a single Nb atom per unit cell.

Given the apparent tunability of the transition temperature with stoichiometry, $Nb_3Cl_4Br_4$ was selected to study the high and low temperature phases with single crystal x-ray diffraction (SXRD). The high temperature phase α-$Nb_3Cl_4Br_4$ undergoes its transition to β-$Nb_3Cl_4Br_4$ at T = 194 K, so diffraction patterns were taken of the same crystal at T = 300 K and T = 110 K. It was found that the high temperature phase was isostructural to α-$Nb_3Cl_8$ (Table S1) and the low temperature structure was isostructural to β-$Nb_3Br_8$ (Table S2). Chlorine and bromine were restricted to occupy identical positions within the structure and their occupancies were allowed to refine with the restraint that each site would remain fully occupied. The structures determined by SXRD for the high and low temperature phases, along with the average site occupancies, can be seen in Figure S1a for the high temperature phase and S1b for the low temperature phase. Highlighted there is the fact that the low temperature phase can be thought of as pairs of layers in the same arrangement as the high temperature phase which are then staggered with each other as to produce a 6 layer unit cell. Labels for the β-phase in Figure 3 are chosen to preserve the layer orientation and arrangement of the α-phase to allow useful comparison. Cross-sectional STEM imaging of the end members, $Nb_3Cl_8$ and $Nb_3Br_8$, directly shows the staggering of the layers in the alpha and beta phase, respectively. In this orientation, the metal atomic columns exhibit two intensity values in STEM, with the bright columns containing more Nb atoms in projection.

Since the SXRD data was taken on a single crystal, it could be used to elucidate some details about the mechanism of the transition. The relationship between the individual halide sites and the $Nb_3$ trimers could be tracked as the halide substitution is not entirely random. The reason for this is believed to be the different potentials for each of the four unique halide sites in the structure, which plays a defining role in the site selectivity of chalcogenide substitution in the $Nb_3QX_7$ (Q = S, Se, Te; X = Cl, Br, I) family of materials.[16] Only one halide site in the structure of the high temperature phase is chlorine rich, which are the intracluster bridging halides, with occupancies provided in table S3. In the beta phase the intracluster bridging halides remain the sole chlorine rich site indicating that the transition must be a mechanical shift of the individual layers and not a change in chemical bonding of individual layers by passing through a kagomé structure from one trimerized kagomé to the other.

**Magnetic Susceptibility.** Previous work[15] has shown negligible orientation dependence on magnetic susceptibility for $Nb_3Cl_8$, therefore samples were prepared as described in the methods section depending on instrument requirements to probe different temperature ranges and the data was combined into Figure 4. Above T = 300 K susceptibility data was collected on oriented single crystals and below T = 300 K data was collected on randomly oriented assemblages of single crystals. Contributions from the sample holder were subtracted out by measuring a blank.

The magnetic susceptibility of the high temperature phase of the series can be approximately fit to a single universal Curie-Weiss curve, $\chi_{f.u.} = C/(T-\theta) - \chi_o$, which can be seen in

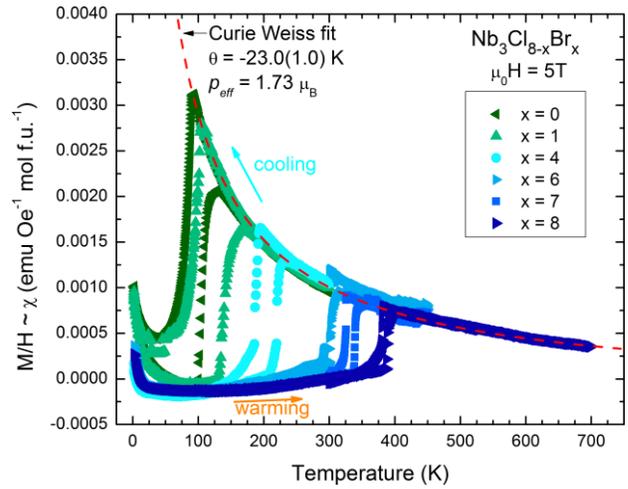

**Figure 4.** Magnetic susceptibility as a function of temperature for $Nb_3Cl_{8-x}Br_x$ samples from x = 0 to x = 8. Hysteresis between warming and cooling can clearly be seen and a universal curve for the paramagnetic regime is included as a dashed red line. This gives for the series a $\theta$ = -23(1) K and $p_{eff}$ = 1.733(9) $\mu_B$ consistent with S = ½ per molecular unit.

Figure 4. This curve for the series average yields a Weiss temperature of $\theta$ = -23.0(1.0) K and a Curie constant $C$ = 0.376(4) emu · K · Oe$^{-1}$ · mol f.u.$^{-1}$, which corresponds to an effective moment, $p_{eff}$ = 1.733(9) $\mu_B$, consistent with $S_{eff}$ = ½. While this universal curve qualitatively does a good job at describing the susceptibility over the entire range of x in $Nb_3Cl_{8-x}Br_x$, it is difficult to tell how closely these values, particularly $\theta$, might conform to the individual fits. As a consequence of the small overall moment, S = ½ per $Nb_3X_8$ unit, significant uncertainties in $\theta$ result from a high sensitivity of the calculated $\theta$ to small changes in $\chi_o$ as the susceptibility becomes nearly linear at higher temperatures. The low temperature phase is accompanied by an almost total loss of magnetization, consistent with the formation of a singlet ground state with a thermal gap to an excited triplet state. The upturn seen in the low temperature regime is attributed to a small number of defect spins from impurities, stacking faults, or edge states which account for between 0.5% and 1.7% of the high temperature spins for all samples. This was determined by fitting the Curie tail. Transition broadening was observed in samples with higher proportions of impurity spins and is typically dependent on many microscopic details including disorder.

For all compounds measured, the transition between a high temperature paramagnetic state and the low temperature state is accompanied by hysteresis, consistent with a first order phase transition, though the hysteresis of the transition is greater for mixed halides than the end members of the $Nb_3Cl_{8-x}Br_x$ family, as shown in table S4. Across the series, the temperature of the transition on warming varies nearly linearly with bromine composition as can be seen in Figure 5a. This demonstrates that the transition can be tuned in a predictable way by varying the stoichiometry of the compound. As previously stated, greater hysteresis is observed for the mixed halides, which could be at least partly attributable to increased supercooling before switching phases due to uneven halide sizes increasing the resistance to structural rearrangement.

**PXRD.** To verify the sample identity, PXRD were taken on a representative sample of each composition produced in this work. Rietveld refinement was not attempted due to the difficulty in addressing the convolution of preferred orientation and stacking defects on the resulting patterns, though lattice parameters were tracked over the compositional range, the results of which can be seen in Figure 5b. For samples in the α-phase at room temperature the 002 and 101 peaks were used to determine values for lattice parameters $c$ and $a$, respectively. For samples in the β-phase at room temperature the 006 and 010 peaks were used to determine the lattice parameters $c$ and $a$ respectively. $Nb_3Cl_2Br_6$, which has a transition above room temperature on warming and below on cooling, was heated to well above its transition temperature prior to collecting PXRD data on it to ensure it would be in its α-phase.

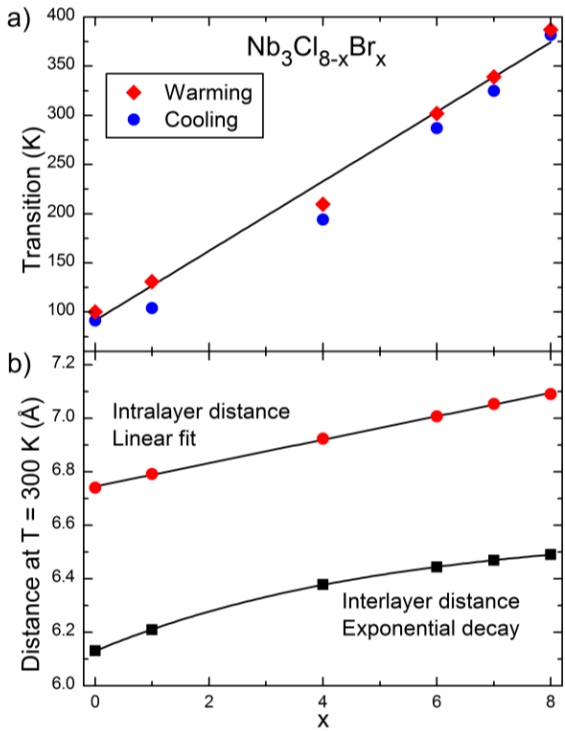

**Figure 5.** a) Transition temperature as a function of stoichiometry, data points mark where the transition occurs on warming (red diamonds) and cooling (blue circles). The temperature of the transition on warming is closer to linear and a line has been provided to guide the eye. b) Distance between the centers of each cluster within a layer is shown by the red circles and follows Vegard's law as shown by the included linear fit, d = 6.745(3) + 0.0436(6) x. The black squares show the average distance between layers which deviates from Vegard's Law and is best fit by an exponential decay, d = 6.585(6) − 0.455(5)$e^{(-x/5.11(12))}$. As the layers are VdW stacked this behavior is not unexpected.

From Figure 5b, it can be seen that the $a$ lattice parameter, which is the same as the distance between the center of each cluster within a layer, varies linearly with substitution of bromine for chlorine. This is in very good agreement with Vegard's law which is an empirical rule that states that, at the same temperature, if the crystal structures of two compounds are the same (which is true for individual layers of $Nb_3Cl_{8-x}Br_x$), and in the absence of any other effects, then a solid solution of those two constituents should result in a lattice parameter that varies linearly between the end members of the series.

The average distance between layers, in the α-phase this is $c/2$ and in the beta $c/6$, deviates considerably from Vegard's law. Since these are VdW layered compounds, this is not unusual since Vegard's law only holds if there are no relevant differences between ions other than their size. Since the layers are not directly bonded to each other, the effective size of ions can change nonlinearly due to differences in polarizability of interacting ions; in particular, bromine is considerably more polarizable than chlorine. The initial bromines substituted for chlorine are constricted to an extent by the surrounding chlorines, which are less polarizable. This leaves the electronically softer bromine to distribute more of its partial charge into the gap between the fairly rigid layers to reduce strain, causing a larger than expected change in interlayer distance than if following Vegard's law. In this case, when a single bromine is substituted per unit cell, the change in $c$ is twice what would be expected. As more bromine is substituted into the structure, both the $a$ lattice parameter and the proportion of Br-Br interactions to Br-Cl interactions increase, resulting in a much smaller change in lattice parameter than would otherwise be expected. The difference between the change in the interlayer and intralayer distances between $Nb_3Cl_8$ and $Nb_3Br_8$ is ultimately less than 0.01 Å

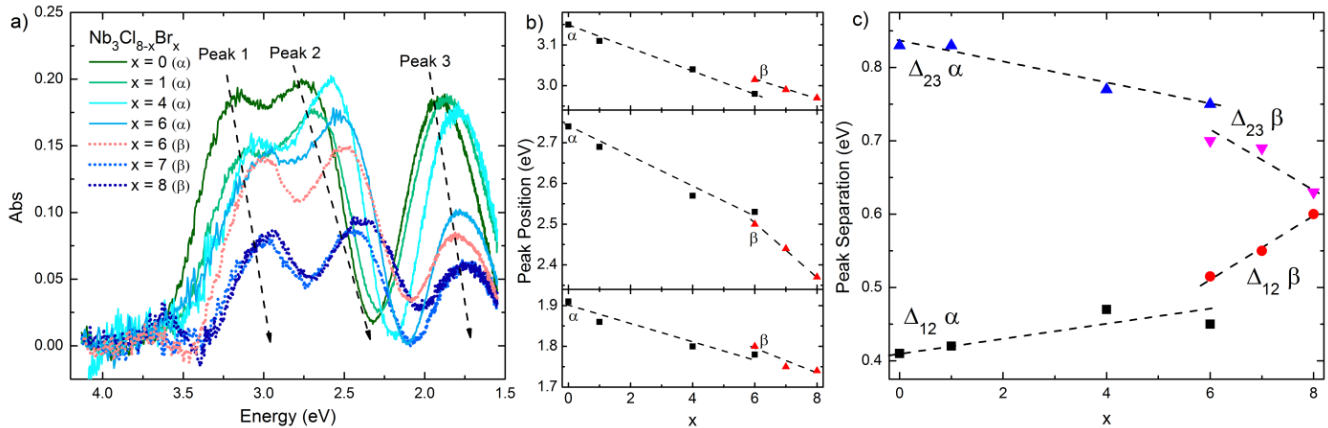

**Figure 6.** a) Absorbance corrected for background and a broad UV absorbance peak for the $Nb_3Cl_{8-x}Br_x$ (x = 0 to x = 8) series. Dashed lines show the general trend of peak locations with increasing x. For $Nb_3Cl_2Br_6$, which has its transition around room temperature, both the α and β phases are shown. b) Absorbance maxima for each of the three peaks in the visible range as a function of stoichiometry. Panels have been sized to provide a consistent scale for all three peaks and lines have been added to guide the eyes. c) Difference in energy of the 3 absorbance peaks in the visible spectrum as a function of stoichiometry with lines added to guide the eyes. Peak 1 is the highest energy peak while peak 3 is the lowest.

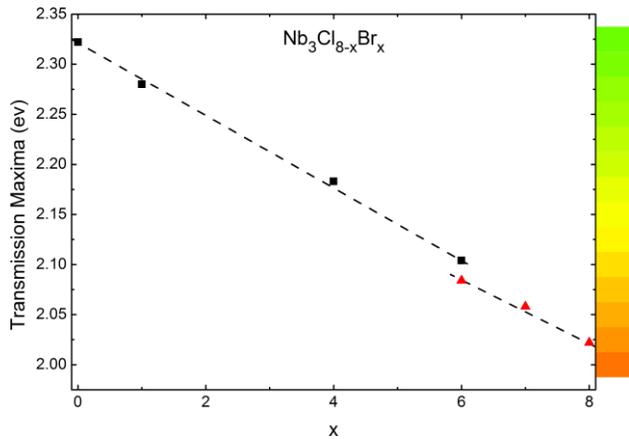

**Figure 7.** Location of the bulk transmission maxima for $Nb_3Cl_{8-x}Br_x$ as a function of stoichiometry. The pure color corresponding to this value is shown in the gradient.

despite the deviations from Vegard's law for the interlayer distance of the intermediate compounds.

**UV-vis spectroscopy.** Crystals were exfoliated using a piece of 3M scotch tape. They were then thinned down by repeated exfoliation to average thicknesses around 200 nm, as determined through cross sectional SEM imaging of a similarly prepared sample on carbon tape. As $Nb_3Cl_2Br_6$ has a transition temperature less than room temperature (cooling) but greater than room temperature (warming), it is possible to stabilize both the high and low temperature forms in a room temperature UV-Vis measurement. This was done by placing the sample assembly in an oven at 50 °C or submerging the sample in liquid nitrogen respectively to switch between the two phases, then allowing the sample to return to room temperature before the measurement.

It was found that there were three absorbance peaks in the visible spectrum, the positions of which, for $Nb_3Cl_8$, are consistent with those previously reported.[17] Peak positions appeared to be insensitive to sample thickness (between ~100 μm to ~100 nm) in experiments involving successive exfoliation. In addition to the three peaks in the visible spectrum the samples had a relatively flat absorbance background along with a broad peak centered in the UV range below the minimum wavelength that could be probed with the setup due to the opacity of 3M tape below 300 nm. To fit this background it was assumed that the UV peak had a Gaussian form and that the three visible peaks would as well if the peak width, position and spectral weight of the UV peak were correctly defined. The UV peak and flat background have been subtracted out in Figure 6a, showing only the contribution from the peaks in the visible spectrum.

As bromine is substituted for chlorine in the structure, all three peaks continuously decrease in energy, which can be seen in Figure 6a and 6b. Upon transitioning from α-$Nb_3Cl_2Br_6$ to β-$Nb_3Cl_2Br_6$ there is a discontinuity for all three peaks, with the first and third peaks jumping to a slightly higher energy by 35 and 20 meV, respectively, and the second peak shifting to lower energies by 30 meV. The highest and lowest energy visible absorbance peaks shift in energy in a very similar way with increasing substitution, suggesting that they may be associated with a transition to or from the same band. The $Nb_3X_8$ family is a multi-element compound with many band crossings at the gamma point making exact identification of the transitions difficult.[17] The middle peak is more sensitive to composition than the other two peaks, which can be seen clearly in Figure 6b as the scale for each panel is kept constant.

After the transition to the β-phase, the rate of change in the energy of the second peak as a function of stoichiometry significantly increases, while the rate of change of the first and third peaks remain effectively constant. The consequences of both the jump in the peak energies and the difference in behavior of the second peak can be seen in Figure 6c. In α-$Nb_3Cl_8$ the ratio $\Delta_{23}/\Delta_{12}$ is 2.02 and in β-$Nb_3Br_8$ it is 1.05 as the peaks become almost equidistant from each other. Across the transition from α-$Nb_3Cl_2Br_6$ to β-$Nb_3Cl_2Br_6$ at room temperature the ratio changes from 1.67 to 1.36 as a result of the alternating direction of the discontinuity across the phase transition. There appeared to be a continuous transformation between the optical absorbance of the high and low temperature phases when scanning in temperature, without major changes in absorptivity.

The actual color of the exfoliated samples appears to be primarily a function of the transmission maxima that occurs between the second and third absorbance peaks. This feature varies highly linearly with stoichiometry, as can be seen in Figure 7. This makes the value a good indicator of the stoichiometry of any given sample. This is useful in the field of device making where very thin samples are used, which may have a stoichiometry different from the average stoichiometry of crystals produced in that batch. Specifically in the context of typical device geometries this transmission maxima would be observed as a reflectance maxima, as can be seen in previous research on $Nb_3Cl_8$ which used diffuse reflectance rather than transmission to study its optical properties.[17]

CONCLUSIONS

$Nb_3Br_8$ was found to undergo the same transition between a high temperature paramagnetic phase and a low temperature singlet ground state observed in $Nb_3Cl_8$. The temperature of this transition can be continuously tuned between that of the two end members by varying the Cl/Br ratio in $Nb_3Cl_{8-x}Br_x$.

There is a significant benefit to any study of the physics in the low temperature phase of this family from the fact that β-$Nb_3Br_8$ appears to share the same magnetic properties as β-$Nb_3Cl_8$. One example is that this allows us to postulate additional constraints on plausible descriptions of the character of the singlets in the low temperature phase which were not accessible when considering $Nb_3Cl_8$ alone. Any such descriptions should also extend to the $Nb_3Cl_{8-x}Br_x$ family as a whole. Previous works have postulated that a symmetry-breaking distortion involving localization of the diffuse $2a_1$ electron,[15] or charge disproportionation between clusters[18] might be responsible for the loss of magnetism observed in β-$Nb_3Cl_8$. Neither explanation seems likely when extended to include β-$Nb_3Br_8$, which, to the resolution limits of SXRD performed on β-$Nb_3Br_8$ here and in prior structural studies[13,19] shows no apparent deviations from the ideal $R\bar{3}m$ structure, in which there is a single unique Nb site. This structure is consistent with that observed in atomic resolution STEM imaging on β-$Nb_3Br_8$. Further work will be required to elucidate the nature of the low temperature singlet ground state in the context of this information.

In the context of devices, the $Nb_3Cl_{8-x}Br_x$ series of compounds possess a number of desirable features. They are 2-dimensional, readily exfoliatable semiconductors with a magnetic phase transition continuously tunable between T = 92 K and T = 387 K. The compounds are effectively air stable in the bulk on the timescale of years. In addition, the series hosts an optical bandgap continuously tunable with composition, with a small discontinuity across the transition between α and β phases when measured at room temperature. The synthesis method described in this paper allows for the rapid growth of large single crystals of any composition in the $Nb_3Cl_{8-x}Br_x$ series suitable for preparation in 2D form. Our work also demonstrates the extreme importance of heterostructure stacking sequence on physical properties in ostensibly 2D VdW layered materials.

METHODS

**Synthesis.** All crystals used in this work were grown through chemical vapor transport. Stoichiometric mixtures of niobium powder (Alfa, 99.99%), $NbCl_5$ (Strem, 99.99%), and $NbBr_5$ (Strem, 99.9%) with a total mass of 1.5 grams were ground together and added to a 14 x 16 mm diameter fused silica tube in a glovebox and handled using standard air free techniques. The tubes were then sealed air free at a length of approximately 30 centimeters, about 5 centimeters longer than the first two zones of a three-zone furnace. For $Nb_3Cl_8$ either 20 mg of $NH_4Cl$ or 40 mg of $TeCl_4$ were added as a transport agent. For $Nb_3Br_8$ either $NH_4Br$ or no transport agent was used with the primary difference being the overall yield of the reaction. The mixed halides were all grown with $NH_4Br$ as the transport agent. A three-zone furnace was used with a temperature gradient of T = 840°C → 785°C → 795°C with all but the last few centimeters of the tube between the first two zones. This discouraged the formation of large intergrown clumps of crystals at the end of the tube. The furnace was held at temperature for 3-5 days before being cooled to room temperature over 7 hours.

Typical larger crystals of $Nb_3Cl_8$ formed with $NH_4Cl$ as the transport agent were thin hexagonal plates with side lengths on the order of 1 cm and masses around 20 mg. Slightly thicker crystals could be obtained when using $TeCl_4$ as the transport agent, up to 40-50 mg. However, energy dispersive x-ray spectroscopy (EDS) indicated the inclusion of tellurium, likely as $Nb_3TeCl_7$. Most $Nb_3Cl_8$ crystals were smaller and growth limited by contact with other crystals being grown on the sides of the tube.

Crystals of $Nb_3Br_8$ are much thicker than those of $Nb_3Cl_8$ with a typical crystal being 2-3 mm thick. Average crystals of $Nb_3Br_8$ were around 60 to 100 mg with the largest crystals reaching masses around ~650 mg. Average crystal size of the intermediate halides appeared to be continuously tuned between that of $Nb_3Cl_8$ and $Nb_3Br_8$ as a function of composition.

**Characterization.** Powder X-ray diffraction (PXRD) data were acquired at room temperature using a Bruker D8 Focus diffractometer with a LynxEye detector using Cu Kα radiation (λ = 1.5424 Å). Lattice parameters were extracted from the data using TOPAZ 4.2 (Bruker).[20]

Single crystal X-ray diffraction (SXRD) data were collected at room temperature and T = 110 K using the program CrysAlisPro (Version 1.171.36.32 Agilent Technologies, 2013) on a SuperNova diffractometer equipped with Atlas detector using graphite-monochromated Mo Kα (λ = 0.71073 Å). CrysAlisPro was also used to refine the cell dimensions and for data reduction. The temperature of the samples was controlled using the internal Oxford Instruments Cryojet. The structures were solved using SHELXS-86 and refined using SHELXL-97,[21] Visualization was done in Vesta.[22]

Magnetic susceptibility data were collected using a Quantum Design Physical Properties Measurement System (PPMS) on either randomly oriented arrays of crystals using the standard option for measurements below T = 300 K, with the exception of $Nb_3Cl_2Br_6$, or perpendicular to the c-axis of single crystals using the Vibrating Sample Magnetometer (VSM) for the T < 300 K measurement of $Nb_3Cl_2Br_6$ and VSM Oven option for all measurements above T = 300 K.

SEM data were collected using a JEOL JSM IT100 scanning electron microscope at 20 keV with EDS option.

Backscattered Laue data were collected using an accelerating voltage of 15 keV and a beam diameter of 0.5 mm.

UV-Vis spectroscopy data were collected on an Agilent Technologies Cary 60 UV-Vis spectrometer.

STEM data was acquired using an aberration-corrected FEI Titan Themis operated at 300 kV. The convergence angle was 21.4 mrad and the collection inner-angle was 68 mrad. For sample preparation, flakes were first exfoliated from a bulk crystal using 3M scotch tape and transferred onto a SiO2/Si substrate. Cross-sectional samples were obtained using focused ion beam liftout. To avoid sample drift artifacts and increase signal to noise, stacks of STEM images were acquired in rapid succession (1 μs/pixel) and aligned using rigid registration methods.[23]

ACKNOWLEDGEMENTS

We acknowledge stimulating discussions with Collin Broholm. We thank Maxime Siegler for assistance with SXRD measurements and analysis. We thank Hector Vivanco for assistance with collecting SEM images. Work at the Institute for Quantum Matter was supported by the U.S. Department of Energy, Office of Basic Energy Sciences, Division of Material Sciences and Engineering under Award No. DE-SC0019331. TMM acknowledges support of the David and Lucile Packard Foundation. STEM measurements were supported by the National Sciences Foundation (NSF) through the Platform for the Accelerated Realization, Analysis, and Discovery of Interface Materials (DMR-1539918). This work made use of the Cornell Center for Materials Research Shared Facilities which are supported through the NSF MRSEC program (DMR-1719875). The FEI Titan Themis 300 was acquired through NSF-MRI-1429155, with additional support from Cornell University, the Weill Institute and the Kavli Institute at Cornell.

# Supporting Information

## Abbreviated refinement tables

**Table S1.** Abbreviated refinement parameters for α-$Nb_3Cl_4Br_4$ refined with anisotropic thermal parameters. Crystallographic information file can be found [html]

| | | | |
|---|---|---|---|
| MW (g/mol) | 740.15 | Z | 2 |
| Crystal System | Trigonal | Temperature | 293 K |
| Space Group | *P-3m1* | λ (Å) | 0.71073 |
| Color | black | Total refl. | 16834 |
| $a = b$ (Å) | 6.9133(2) | Unique refl. | 930 |
| $c$ (Å) | 12.7267(3) | Parameters | 29 |
| α (°) | 90 | $R_{int}$ | 0.0393 |
| β (°) | 90 | GooF | 1.289 |
| γ (°) | 120 | $R(F_o)$ | 0.0164 |

**Table S2.** Abbreviated refinement parameters for β-$Nb_3Cl_4Br_4$ refined with anisotropic thermal parameters. Crystallographic information file can be found [html]

| | | | |
|---|---|---|---|
| MW (g/mol) | 740.15 | Z | 6 |
| Crystal System | Trigonal | Temperature | 110 K |
| Space Group | *R-3m* | λ (Å) | 0.71073 |
| Color | black | Total refl. | 3289 |
| $a = b$ (Å) | 6.9034(4) | Unique refl. | 222 |
| $c$ (Å) | 38.035(2) | Parameters | 29 |
| α (°) | 90 | $R_{int}$ | 0.0290 |
| β (°) | 90 | GooF | 1.253 |
| γ (°) | 120 | $R(F_o)$ | 0.0121 |

## Structures from SXRD

## Site occupancies

There are 4 unique halide sites for both the α and β phases of the $Nb_3X_8$ family of materials. Two of the sites are replicated only once per formula unit and are the intracluster and intercluster capping halides. The other two sites are replicated three times per formula unit and each serve to bridge a pair of Nb atoms. These sites are the intracluster and intercluster bridging halides. The occupancies found for these sites in the mixed halides serve as unique chemical tag in single crystal x-ray diffraction, allowing a single sample to be tracked above and below its transition temperature to provide some restraints on the mechanisms by which the transition can occur. The small discrepancies between α and β are likely due in part to the rough absorption correction on the β phase data and consequences of the stacking faults seen in the low temperature phase.

**Table S3.** Bromine site occupancies determined through single crystal x ray diffraction for α and β $Nb_3Cl_4Br_4$

| Phase | α-$Nb_3Cl_4Br_4$ | β-$Nb_3Cl_4Br_4$ |
|---|---|---|
| Intracluster cap | 0.6485 | 0.6462 |
| Intercluster cap | 0.6617 | 0.6580 |
| Intercluster bridge | 0.5021 | 0.5035 |
| Intracluster bridge | 0.3975 | 0.4020 |

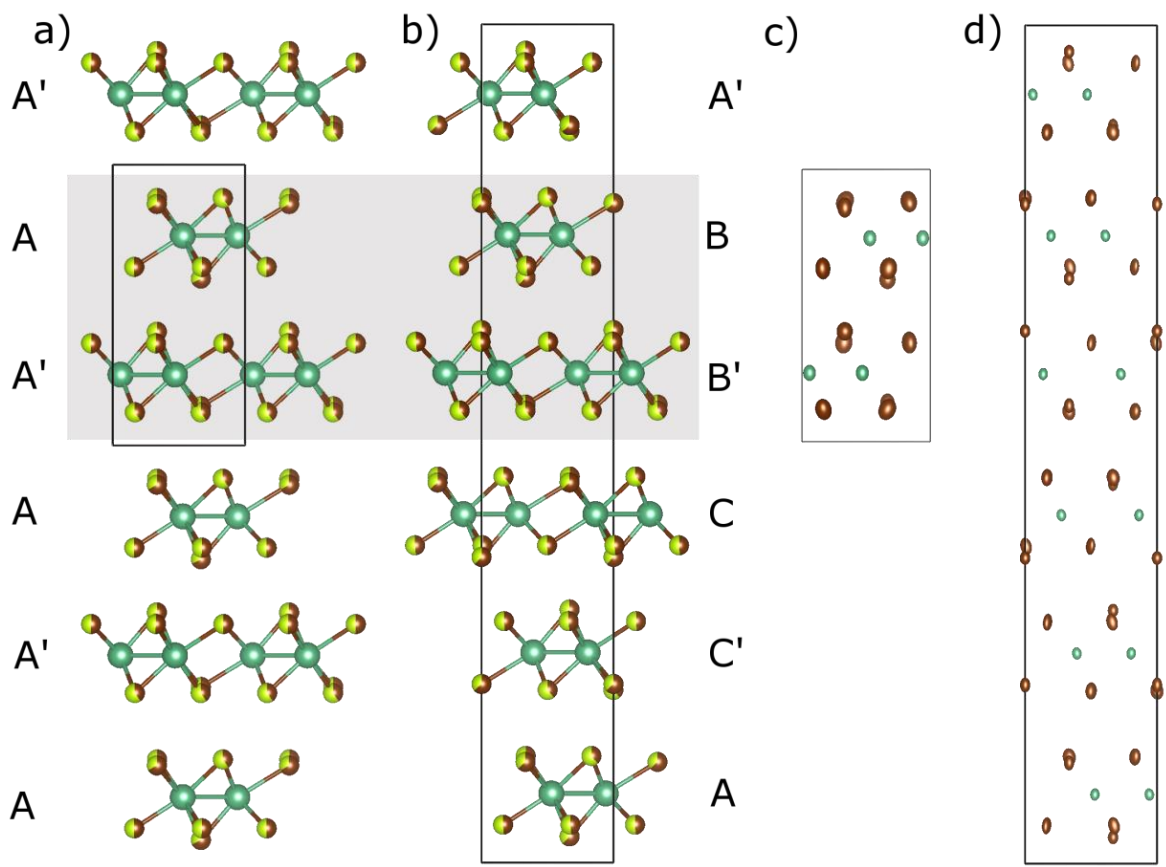

**Figure S1.** a) High temperature alpha phase showing AA' stacking of $Nb_3Cl_4Br_4$. b) Low temperature beta phase showing AA'BB'CC' stacking of $Nb_3Cl_4Br_4$. The highlighted pair of layers show how the β phase consists of staggered pairs of layers related to each other by the AA' stacking seen in the α phase, also demonstrated by the labeling scheme used. Green represents niobium, yellow is chlorine and brown is bromine. c) and d) show the unit cells with anisotropic thermal parameters for the α and β phases respectively. The stoichiometry for the particular sample used, determined from the high temperature SXRD was $Nb_3Cl_{3.99}Br_{4.01}$.

**Hysteresis and Transition broadening**

Observed in Figure 4 is both hysteresis and transition broadening. The hysteresis is defined by the difference between the temperatures at which the transition begins on warming and cooling. This is consistent with what is expected for a first order phase transition. The increased hysteresis in samples with more disorder (mixed halides), especially at transitions that occur at lower temperatures, is also consistent with the apparent mechanism of the transition as some form of mechanical shift rather than a rearrangement of chemical bonds through a breathing mode. The exact mechanism by which this occurs, especially when considering the low temperatures at which it happens in samples with high amounts of chlorine, is still being investigated.

The broadening is the temperature range over which the transition occurs once it does begin. Two samples show significant transition broadening, $Nb_3Cl_8$ and $Nb_3Cl_7Br$, The broadening is sample dependent and thus likely arises from details of the defects and disorder present[15]. All samples in which a single crystal were measured instead of an assembly of crystals show less broadening, suggesting that inconsistencies between individual crystals also plays a role in aggregate measurements.

**Table S4.** Hysteresis as a function of stoichiometry for $Nb_3Cl_{8-x}Br_x$

| Stoichiometry | Hysteresis (K) |
| --- | --- |
| x = 0 | 8.5 K |
| x = 1 | 27 K |
| x = 4 | 15.5 K |
| x = 6 | 15 K |
| x = 7 | 14 K |
| x = 8 | 5 K |